\documentclass[aps,prl,floatfix,twocolumn]{revtex4}
\usepackage{amsmath,amssymb}
\usepackage{graphicx}
\usepackage{epsfig}
\usepackage{psfrag}
\usepackage{color}
\usepackage[dvipsnames]{xcolor}

\newcommand{\bm}{\boldsymbol}

\begin{document}

\hsize\textwidth\columnwidth\hsize\csname@twocolumnfalse\endcsname

\title{
Magneto-Coulomb Drag and Hall Drag 
in Double-Layer Dirac Systems}

\author{Wang-Kong Tse$^1$}
\author{B. Y.~K. Hu$^2$}
\author{J.~N. Hong$^1$}
\author{A.~H. MacDonald$^3$}
\affiliation{$^1$Department of Physics and Astronomy, Center for
Materials for Information Technology, The University of Alabama, Alabama 35487, USA}
\affiliation{$^2$Department of Physics, The University of Akron, Akron,OH 44325, USA}
\affiliation{$^3$Department of Physics, University of Texas, Austin, Texas
  78712, USA}

\begin{abstract}
We develop a theory of Coulomb drag due to momentum transfer 
between graphene layers in a strong magnetic field. 
The theory is intended to apply in systems with disorder that is weak compared to Landau level separation, so
that Landau level mixing is weak, but strong compared to correlation energies within a single Landau level, so that 
fractional quantum Hall physics is not relevant.
We find that in contrast to the zero-field limit,
the longitudinal magneto-Coulomb drag
is finite, and in fact attains a maximum at the simultaneous charge 
neutrality point (CNP) of both layers.  Our theory also predicts a sizable Hall drag
resistivity at densities away from the CNP.  
\end{abstract}

\maketitle

\noindent
\textit{Introduction.---} Progress in preparing high quality samples of 
graphene \cite{GraphRev1,GraphRev2} and other 
atomically thin two-dimensional (2D) systems has made it possible to study interlayer
interaction effects in Coulomb-coupled electron gas
layers separated by only a few nanometers.  The archetypical Coulomb-coupling phenomenon
is drag resistance due to Coulomb interactions between layers \cite{DragRMP}.
When an external electric field drives a current through one of the
layers, there is a non-zero rate of net 
momentum transfer from electrons in the drive layer to 
electrons in the drag layer, resulting in a {\it drag voltage} in an open-circuit geometry. 
This intriguing effect has been extensively studied theoretically and
experimentally in conventional 2D electron gas (2DEG) systems.
Atomically thin Dirac electron systems like graphene present new 
challenges to theories of Coulomb drag because stronger coupling
can be achieved by placing the two layers closer together \cite{vdWH1,vdWH2}.

The weak coupling regime of Coulomb drag in double-layer graphene
system has been explored theoretically \cite{Tse_GDrag,Dragpaper1,Dragpaper2,Dragpaper3,Dragpaper4,Dragpaper5,Dragpaper6} and realized experimentally \cite{Tutuc_exp1}.
Striking differences compared with the zero field case
are observed \cite{Geim_drag1,Geim_drag2,Dean_exp1,Tutuc_exp2} 
when closely separated ($\gtrsim 1\,\mathrm{nm}$) double-layer graphene or bilayer
graphene structures are placed in weak perpendicular magnetic fields. 
The observations of a finite longitudinal drag resistivity at
the simultaneous CNP of both layers and of a large Hall drag away from
this density have been particularly intriguing.  
Very recently, strong-field Coulomb drag in the quantum
Hall regime has been measured in double layers of 
graphene \cite{Kim_exp2} and bilayer 
graphene \cite{Dean_exp2,Kim_exp}. 

In this Letter, we develop a theory of Coulomb 
drag due to \textit{momentum exchange} between 2D graphene sheets
in the presence of strong magnetic fields.
We consider the case where fields are strong enough for weakly disorder-broadened 
Landau levels to be well-resolved.  Our theory 
does not apply when an exciton condensate is present, or in the 
weak-disorder, low-temperature regime at which the fractional quantum 
Hall effect and other phenomena associated with strong electronic correlations appear.
We find that the drag resistivity behaves very differently 
in the strong magnetic field and zero magnetic field limits. 
As we will show, a finite drag resistivity is present at the
simultaneous CNP of both layers at strong fields, whereas 
drag vanishes at that point in the $B=0$ case \cite{Tse_GDrag}. 
Away from the CNP, we find a sizable
Hall drag resistivity. These two main findings from our theory 
corroborate recent experimental observations \cite{Kim_exp2}.

\noindent
\textit{Theory.---}  The eigenstates of the  graphene free-particle Hamiltonian consist
of a special Dirac point Landau level (LL) labelled by $(n=0,X)$, 
and other LLs labeled by $(n,s,X)$, where $n = 1, 2, \dots$ 
is a LL index, $s = \pm$ is a band label (conduction band $+$ and valence
band $-$),
and $X$ is a Landau gauge guiding center label.
The LL energies are $s\varepsilon_{n} = s\hbar\omega_{\mathrm{c}}\sqrt{n}$ where
$\omega_{\mathrm{c}} = {\sqrt{2}v}/{\ell_B}$ with $v$ the Dirac 
velocity ($v = 10^6\,\mathrm{ms}^{-1}$ in graphene) and $\ell_B = \sqrt{\hbar/e\vert
  B\vert}$ the magnetic length.
\begin{figure}
\includegraphics[width=8.5cm,angle=0]{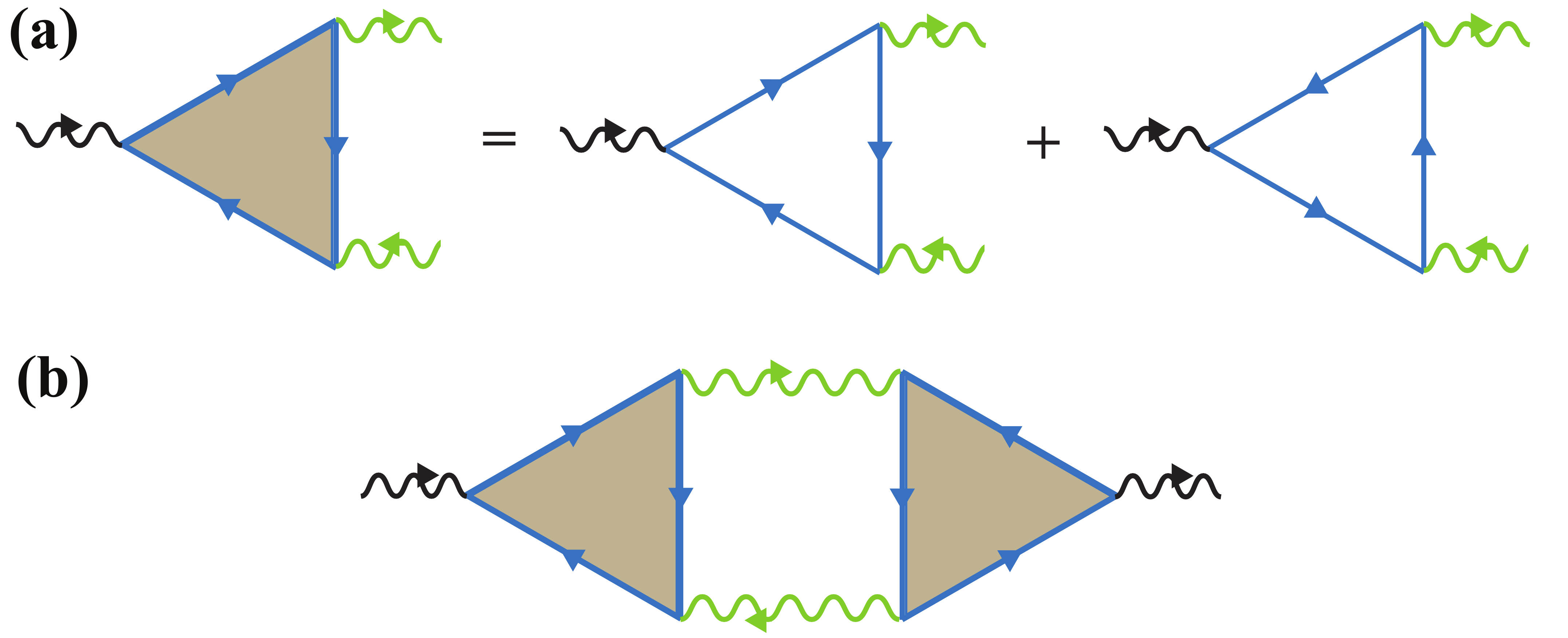}
\caption{(Color online) (a) Diagrams for the nonlinear susceptibility
$\Gamma$. The black (dark) wavy line denotes a vector potential
coupled to a current vertex and the light (green) wavy lines
denote scalar potentials coupled charge vertices. (b) Diagram for
the drag transconductivity.  Interlayer screened Coulomb interactions ( green (light)
wave lines representing interlayer Coulomb interactions $U(q,\omega)$
link the two triangle nonlinear susceptibility diagrams.} \label{Gamma_Drag}
\end{figure}
The central quantity in the Coulomb drag problem \cite{Oreg,Jauho} is the nonlinear
susceptbility $\bm{\Gamma}(\bm{q},\omega,B)$ [See Fig.~\ref{Gamma_Drag}(a)].
We now derive an expression for this quantity that is valid
for graphene in a magnetic field.
%
%
%
The Green's function in Fig.~\ref{Gamma_Drag}(a) is 
\begin{equation}
G_{n,s,X}(\varepsilon) =  \frac{\vert n,s,X \rangle \langle n,s,X \vert}{\varepsilon-\varepsilon_{s,n} + i/2\tau},
\label{Green}
\end{equation}
where $\vert n,s,X \rangle$ is an eigenstate \cite{eigen_remark} and $\tau$ is its lifetime.
The three vertices in the nonlinear susceptibility diagram Fig.~\ref{Gamma_Drag}(a) contain matrix
elements of the current and charge density operators between different LL wavefunctions.  In a 
continuum model, the eigenstates are spinors with components on both honeycomb sublattices.  
The current and charge density matrix elements are:
%
%
\begin{eqnarray}
&&\langle n_2,s_2,X_2 \vert j_x \vert n_3,s_3,X_3
\rangle = ev \delta_{X_2,X_3} \nonumber \\
&&\times \mathcal{N}_{n_2}\mathcal{N}_{n_3}\left[s_2\delta_{n_2-1,n_3}
+s_3\delta_{n_2,n_3-1}
\right], \label{jmat} \\
&&\langle n_1,s_1,X_1 \vert e^{i\bm{q}\cdot\bm{r}} \vert n_2,s_2,X_2
\rangle = \delta_{X_1,X_2+q_y\ell_B^2}e^{iq_x(X_1+X_2)/2} \nonumber \\
&&\times \mathcal{N}_{n_1}\mathcal{N}_{n_2}\left[F_{n_1,n_2}(\bm{q})+s_1s_2F_{n_1-1,n_2-1}(\bm{q})\right], \label{cmat}
\end{eqnarray}
where $\mathcal{N}_{n} = \delta_{n,0}+(1-\delta_{n,0})/\sqrt{2}$ 
is a normalization factor, and 
we have defined
\begin{equation}
F_{n_1,n_2}(\bm{q}) =
\sqrt{\frac{n_{<}!}{n_{>}!}} e^{-q^2\ell_B^2/4}
L_{n_{<}}^{n_{>}-n_{<}}\left(\frac{q^2\ell_B^2}{2}\right) \left(\frac{i\tilde{q}\ell_B}{\sqrt{2}}\right)^{n_{>}-n_{<}}.
\label{Fq}
\end{equation}
%
In Eq.~(\ref{cmat}), $q = \vert \bm{q} \vert$, $\tilde{q} = q_x+iq_y$,
$n_{>} = \mathrm{max}(n_1,n_2)$, $n_{<} = \mathrm{min}(n_1,n_2)$, 
and $L_n^{\alpha}$ is the generalized Laguarre polynomial of degree $n$.
Evaluating Fig.~\ref{Gamma_Drag}(a) using the standard Matsubara Feynman diagram 
technique and Eqs.~(\ref{Green})-(\ref{Fq}), we obtain the following compact expression 
for the nonlinear susceptibility: 
%
\begin{equation}
\bm{\Gamma}(\bm{q},\omega,\bm{B}) =
2\ell_B^2\bm{q}\times\hat{\bm{B}}\,\mathrm{Im} \; \Pi({q},\omega,B), \label{Gamma}
\end{equation}
where $\hat{\bm{B}}$ is the direction of the magnetic field, 
\begin{eqnarray}
&&\Pi({q},\omega,B) = -\frac{g}{2\pi \ell_B^2} \label{Pi} \sum_{n_1,n_2 =
  0}^{\infty}\sum_{s_1,s_2 = \pm}
\\
&&\frac{f_{s_1,n_1}-f_{s_2,n_2}}{\omega+s_1\varepsilon_{n_1}-s_2\varepsilon_{n_2}+i/2\tau}
\; \; \mathcal{F}_{s_1,s_2}(q\ell_B,n_1,n_2), \nonumber
\end{eqnarray}
$g = 4$ accounts for the spin and valley degeneracy, and $f_{s,n}$
is the Fermi occupation factor for the $(n,s,X)$ LL.
The form factor $\mathcal{F}$ in Eq.~(\ref{Pi}) is 
%
\begin{eqnarray}
&&\mathcal{F}_{s_1,s_2}(x,n_1,n_2)
=
\frac{e^{-x^2/2}}{4}\left(\frac{x^2}{2}\right)^{n_{>}-n_{<}}
\nonumber \\
&&\times \left[s_1\sqrt{\frac{n_{<}!}{n_{>}!}}L^{n_{>}-n_{<}}_{n_{<}}\left(\frac{x^2}{2}\right)\right.\nonumber \\
&&\left.+s_2\sqrt{\frac{(n_{<}-1)!}{(n_{>}-1)!}}L^{n_{>}-n_{<}}_{n_{<}-1}\left(\frac{x^2}{2}\right)\theta\left(n_{<}-1\right)
\right]^2,
\label{matelm}
\end{eqnarray}
where 
$\theta(x) = 1$ for $x \geq 0$ and $0$ otherwise. The quantity $\Pi$ in Eq.~(\ref{Pi})
is the polarization function of Dirac fermions in a perpendicular quantizing magnetic field \cite{Polarpapers}.  
The nonlinear susceptibility $\bm{\Gamma}$ is therefore directly proportional to the 
imaginary part of the polarization function, as in the conventional 2DEG case
with a single \textit{non-chiral} parabolic band
\cite{Bonsager}. 

This finding might seem surprising since the $\Gamma(q,\omega)
\propto \mathrm{Im}\,\Pi(q,\omega)$ property of a conventional 2DEG \cite{Oreg,Jauho}
does not apply to the nonlinear susceptibility of graphene \cite{Dragpaper1} in the absence 
of a magnetic field. 
This difference can be explained by noting that while the
energy dispersions of the conventional 2DEG and
graphene are different at $B = 0$, both have dispersionless Landau levels
at strong $B$. We therefore conjecture 
that, in a strong magnetic field when disorder does not appreciably
mix Landau levels, the simple relationship $\Gamma \propto \mathrm{Im}\,\Pi(q,\omega)$ is a
universal feature of \textit{all} clean two-dimensional electron systems, 
regardless of their energy dispersions. In such a case the nonlinear susceptibility is,
like the polarization function, dominated by inelastic inter-LL transitions of 
electrons from one localized LL orbit to another localized LL orbit. 

Another remarkable distinction of the strong magnetic field is brought
to light by examining the drag resistivity at the CNP.
The nonlinear susceptibility of
graphene in the absence of a magnetic field was first evaluated
in Ref.~\cite{Tse_GDrag}. Making use of the electron-hole symmetry of the bands 
and time-reversal invariance it is straightforward to show that, at
$B=0$, $\bm{\Gamma}({q},\omega)$ is an odd function of the chemical potential $\mu$.
When the chemical potential is at the Dirac point, $\Gamma = 0$ because the two
diagrams comprising Fig.~\ref{Gamma_Drag}(a) exactly cancel. 
Drag therefore vanishes when either layer is charge neutral. At high temperatures 
this behavior has indeed been observed
experimentally \cite{Tutuc_exp1,Geim_drag1}. 
In the presence of a strong magnetic field, on the other hand, the nonlinear susceptibility Eq.~(\ref{Gamma})
is an \textit{even} function of $\mu$, as we prove below. 

First, we note that electron-hole symmetry
is preserved for the $s = \pm$ LLs so that $f_{s,n}(\mu) = 1-f_{-s,n}(-\mu)$, and that   
the form factor in Eq.~(\ref{matelm}) is invariant under 
$s_1 \to -s_1$ and $s_2 \to -s_2$.  Next we can interchange the labels $n_1
\leftrightarrow n_2$, $s_1 \leftrightarrow s_2$ to conclude that 
$\Pi({q},\omega,B)$ is invariant under $\mu \to - \mu$.
It then follows from Eq.~(\ref{Gamma}) that
$\bm{\Gamma}(\bm{q},\omega,\bm{B};\mu) =
-\bm{\Gamma}(-\bm{q},\omega,\bm{B};-\mu)=
\bm{\Gamma}(\bm{q},\omega,\bm{B};-\mu)$.
The final identity requires the observation that the form factors in 
Eq.~(\ref{matelm}) depend on $q = \vert\bm{q}\vert$ only.
Therefore, the contributions from the two diagrams in Fig.~\ref{Gamma_Drag}(a)
do not cancel at $\mu = 0$ as they do in the absence of broken
time-reversal symmetry. 
Drag can
be finite even when one of the layers is charge neutral. 



The interlayer transconductivity diagrams \cite{Oreg,Jauho} yield the drag 
conductivity [Fig.~\ref{Gamma_Drag}(b)]
\begin{eqnarray}
&&\sigma_{\alpha\beta}^{\mathrm{D}} = \frac{e^2}{16\pi\hbar k_{\mathrm{B}}T}
\sum_q\int_{-\infty}^{+\infty}\frac{\mathrm{d}\omega}{\mathrm{sinh}^2(\omega/2k_{\mathrm{B}}T)}
\nonumber \\
 &&\times {\Gamma}_{\alpha}^{\mathrm{L}}(\bm{q},\omega,\bm{B}){\Gamma}_{\beta}^{\mathrm{R}}(\bm{q},\omega,-\bm{B})\vert
 U(q,\omega,B)\vert^2,
 \label{Dragcond}
 \end{eqnarray}
where the superscripts `L' and `R' (left and right) label the two
layers, and $U(q,\omega,B)$ is the screened interlayer Coulomb
interaction in the random phase approximation \cite{RPA}. 
The Coulomb interaction strength in graphene is characterized by the 
dimensionless coupling constant $\alpha_{\mathrm{G}} = e^2/(\epsilon
\hbar v)$, where $\epsilon$ is an effective dielectric constant which we view as a parameter
that can be altered by changing the sheet's dielectric environment \cite{USupp}. 
The quantity measured in most Coulomb
drag experiments is the drag resistivity, which can be obtained by
inverting the four component (two layers each with two directions)
conductivity tensor $\overleftrightarrow{\sigma}$ of the
bilayer , $\overleftrightarrow{\rho} = (\overleftrightarrow{\sigma})^{-1}$.
The conductivity tensor becomes diagonal in Cartesian labels when
$\hat{x}$ and $\hat{y}$ components are replaced by 
left and right handed components ($\hat{x} \pm i \hat{y}$) components.
It simplifies further in the special case of identical left and right sheets
since parallel flow and counterflow are then decoupled.
 
For the general case we introduce the definitions:
%
\begin{eqnarray}
\mathcal{S}_{xx} &=&
{(\sigma_{xx}^{\mathrm{D}})^2-\sigma_{xx}^{\mathrm{L}}\sigma_{xx}^{\mathrm{R}}+\sigma_{xy}^{\mathrm{L}}\sigma_{xy}^{\mathrm{R}}}, \label{Sxx} \\
\mathcal{S}_{xy}
&=&
{\sigma_{xy}^{\mathrm{L}}\sigma_{xx}^{\mathrm{R}}+\sigma_{xx}^{\mathrm{L}}\sigma_{xy}^{\mathrm{R}}}, \label{Sxy}
%
\end{eqnarray}
where $\sigma_{xx}^{\mathrm{L,R}}$ and $\sigma_{xy}^{\mathrm{L,R}}$
are the longitudinal and Hall conductivities in the individual
layers.  Because the Hall drag conductivity $\sigma_{xy}^{\mathrm{D}}$
vanishes due to the odd momentum dependence of $\bm{\Gamma} \propto
\bm{q}\times\hat{\bm{B}}$ in Eq.~(\ref{Gamma}),
the general drag resistivity tensor expression simplifies to
\begin{equation}
\rho_{\alpha\beta}^{\mathrm{D}} =
\sigma_{xx}^{\mathrm{D}}{\mathcal{S}_{\alpha\beta}}/({\mathcal{S}_{xx}^2+\mathcal{S}_{xy}^2}), \label{rhoDS}
\end{equation}
where $\alpha,\beta = x$ or $y$. In 2DEG systems, 
the Hall drag resistivity is negligible for $\omega_{\mathrm{c}}\tau \ll 1$, where $\omega_{\mathrm{c}}$ is
the cyclotron frequency \cite{Oreg,Hu_Hall}. In strong magnetic
fields,  
the Hall drag resistivity is finite and can be significant, arising
from the longitudinal drag combined 
with the intralayer Hall responses $\sigma_{xy}^{\mathrm{L,R}}$.
From Eq.~(\ref{Sxy}), we observe that the Hall drag $\rho_{xy}^{\mathrm{D}}$
is comparable to the longitudinal drag $\rho_{xx}^{\mathrm{D}}$ in magnitude
except when (1) both layers are characterized by
well-formed quantum Hall plateaus such that the longitudinal
conductivities vanish $\sigma_{xx}^{\mathrm{L,R}} = 0$; or (2), 
 since $\sigma^{L,R}_{xx}(\mu)=\sigma^{L,R}_{xx}(-\mu)$ and $\sigma_{xy}^{L,R}(\mu)=-\sigma_{xy}^{L,R}(-\mu)$,
 the two layers have opposite carrier densities \cite{Geim_drag1,Geim_drag2,Tutuc_exp1}.

\begin{figure}
\includegraphics[width=8.5cm,angle=0]
{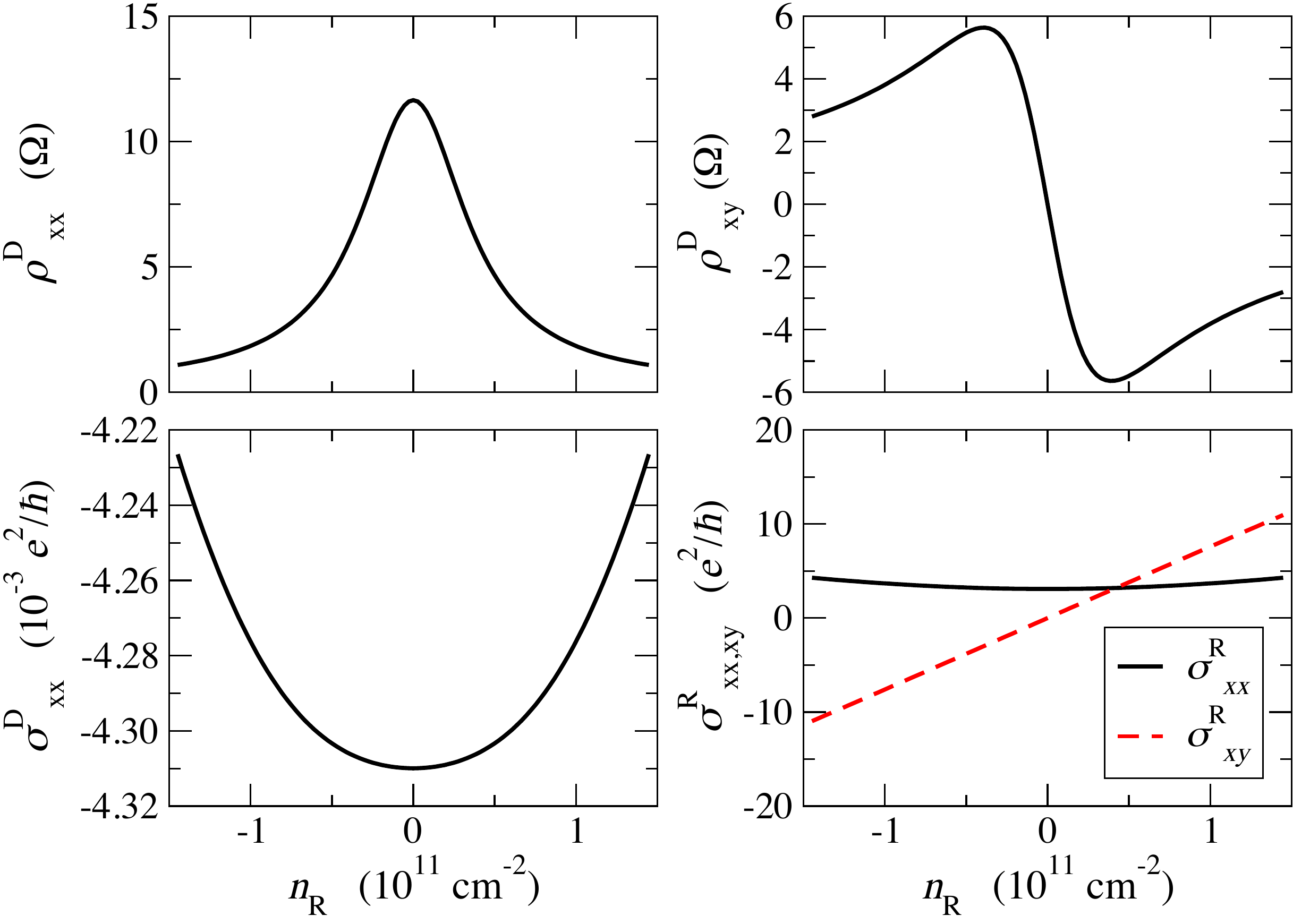}
\caption{(Color online)
Longitudinal (a) and Hall (b) drag
 resistivities of double-layer graphene as a function of right-layer electron density $n_{\mathrm{R}}$ for
$n_{\mathrm{L}} = 0$,  $B = 0.5\,\mathrm{T}$, $T = 300\,\mathrm{K}$, $d = 30\,\AA$,
$1/2\tau = 1\,\mathrm{meV}$, and  $\alpha_{\mathrm{G}} = 0.4$. 
Panel (c) shows the corresponding drag conductivities
and (d) the longitudinal and Hall conductivities.}
 \label{rhoxxxy_0.5_n}
\end{figure}


\noindent
\textit{Dirac-Point Drag and Hall Drag.---}
In the following we present
numerical results for the drag resistivities evaluated from
Eqs.~(\ref{Dragcond})-(\ref{rhoDS}).
We employ the Thomas-Fermi 
approximation in the screened interlayer Coulomb interaction \cite{USupp}. 
%
We first keep the density of one layer ($n_L$) fixed at the CNP and 
vary the density of the other layer ($n_R$).
Fig.~\ref{rhoxxxy_0.5_n}(a) shows the longitudinal drag resistivity as a function of density
in the vicinity of the CNP for $B =0.5\,\mathrm{T}$. 
The most important feature we find is that $\rho_{xx}^{\mathrm{D}}$
has its maximum value at the simultaneous CNP $n_{\mathrm{L,R}} = 0$. Away from the CNP,
$\rho_{xx}^{\mathrm{D}}$ is an even function of $n_{R}$ and decreases with its magnitude.
In Fig.~\ref{rhoxxxy_0.5_n}(b) we show the Hall drag
resistivity $\rho_{xy}^{\mathrm{D}}$, which is an odd function of $n_{R}$.
The magnitude of $\rho_{xy}^{\mathrm{D}}$ rises sharply from zero away
from the simultaneous CNP and then drops gradually as the 
layer's carrier density 
is further increased. 
In Fig.~\ref{rhoxxxy_0.5_n}(c)-(d) we also depict the behavior of the drag
conductivity as well as the R layer's longitudinal and Hall conductivities.
As shown in Fig.~\ref{rhoxxxy_0.5_n}(c), the magnitude of the drag conductivity $\vert\sigma_{xx}^{\mathrm{D}}\vert$ decreases with
density. We note that the sign of the drag conductivity 
is negative and its value is three orders 
of magnitude smaller than the longitudinal and Hall conductivities
$\sigma_{xx,xy}$, which results in a positive sign of the drag resistivity
$\rho_{xx}^{\mathrm{D}}$. 

\begin{figure}
\includegraphics[width=8.5cm,angle=0]
{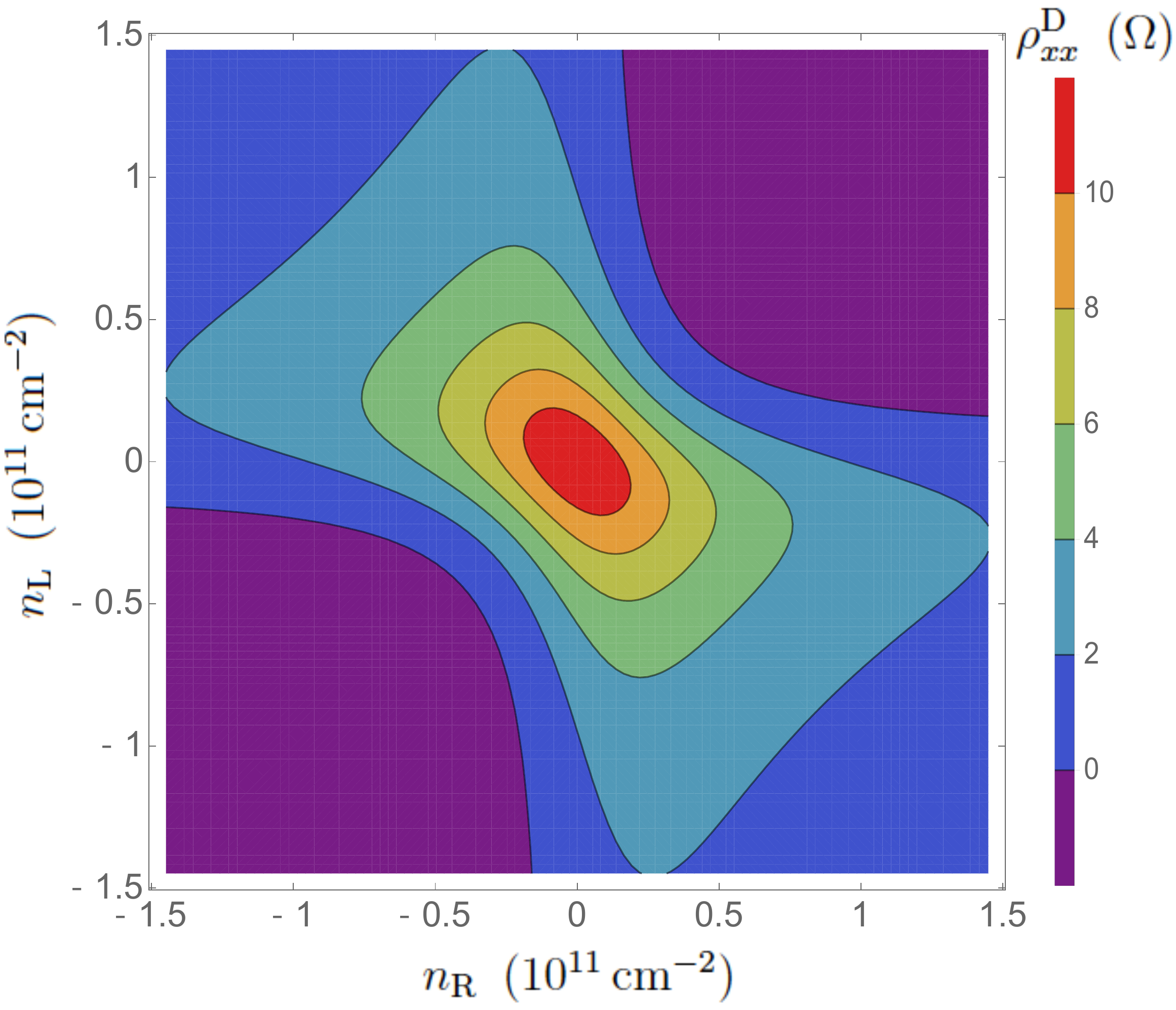}
\caption{(Color online)
Longitudinal drag resistivity as a function of right-layer electron
density $n_{\mathrm{R}}$ and left-layer electron
density $n_{\mathrm{L}}$ for the same values of $B, T, d$, and
$\alpha_{\mathrm{G}}$ as in Fig.~\ref{rhoxxxy_0.5_n}.}
 \label{rhoxx_0.5_3D}
\end{figure}

Fig.~\ref{rhoxx_0.5_3D} shows the dependence of
$\rho_{xx}^{\mathrm{D}}$ on $n_{\mathrm{L}}$ and
$n_{\mathrm{R}}$. We observe that $\rho_{xx}^{\mathrm{D}}$ is positive
in the two quadrants of electron-hole drag where the `L' and `R' carriers
have opposite polarities, and mostly negative  
in the quadrants of electron-electron and hole-hole
drag, with $\rho_{xx}^{\mathrm{D}} < 0$ except near the simultaneous
CNP at $n_{\mathrm{L,R}}\lesssim 10^{10}\,\mathrm{cm}^{-2}$. These
features 
are in good agreement with the latest
experiment \cite{Kim_exp2} performed under strong magnetic fields. 
With the exception of the CNP vicinity, the signs of 
magnetodrag  for the cases of same and opposite carrier polarities depicted in Fig.~\ref{rhoxx_0.5_3D} are
the same as that in the zero-field case 
\cite{Tse_GDrag,Dragpaper1,Tutuc_exp1,Geim_drag1} and consistent with
magnetodrag in conventional 2DEG \cite{Bonsager}. Unlike conventional
2DEG however, which would exhibit no drag when the carrier density is tuned to zero in one of the
layers, double-layer graphene exhibits a distinctive finite magnetodrag at the
simultaneous CNP due to the presence of a Dirac sea of electrons and a 
gapless energy dispersion.
While interlayer energy transfer has been proposed as a possible
mechanism \cite{Dragpaper6} to explain the finite drag
resistivity at the CNP observed in zero-field experiments, we emphasize that such
a mechanism is not necessary to explain the finite positive drag at CNP in the 
presence of a magnetic field, as our findings have demonstrated. 


The fact that $\sigma^D_{xx}$ is non-zero and has a negative sign at
$\mu=0$ can be physically  explained as follows.  Let us assume that the magnetic field is in the $z$-direction and the electric field is applied to the active layer in the positive $y$-direction.  Assuming that the longitudinal components are negligible compared to the transverse components of the intralayer $\sigma^{\mathrm{L,R}}$, this implies that particle currents in the active layer are in the positive $x$-direction (independent of whether they are electrons or holes).  Therefore, the drag force on the passive layer is in the positive $x$-direction.   This acts like an effective electric field in the positive (negative) $x$-direction for holes (electrons), which results in an ``$\mathbf E$"$\times \mathbf B$ drift of the holes (electrons) in the negative (positive) $y$-direction.  Hence, for both electrons and holes, the electric current in the passive layer is in the negative $y$-direction; {\em i.e.}, $\sigma^{D}_{xx} < 0$.
%
At finite temperature the drag currents due to thermally excited electrons and holes reinforce
each other and do not cancel.   The drag
conductivity is an even function of the chemical potential, 
consistent with the evenness of $\Gamma$ discussed above. 

Since a negative longitudinal conductivity results in a thermodynamic instability, 
our findings beg an important question: does a negative drag
conductivity also implies a thermodynamic instability? The condition for
thermodynamic stability is that the conductivity matrix be positive definite.  
In the presence of a magnetic field, the Hall conductivities render
the conductivity matrix $\overleftrightarrow{\sigma}$ anti-symmetric. 
For an arbitrary square matrix that is not necessarily symmetric, the
positive definiteness condition depends on the positivity of the
determinant of the symmetric part of the matrix only \cite{Johnson}.
It follows that the Hall conductivities $\sigma_{xy}^{\mathrm{L,R}}$ drop 
out and the resulting determinant is given by
$\sigma_{xx}^{\mathrm{L}}\sigma_{xx}^{\mathrm{R}}-(\sigma_{xx}^{\mathrm{D}})^2$,
which is positive definite, sharing the same expression with the $B =
0$ case. 


\begin{figure}
\includegraphics[width=8.5cm,angle=0]
{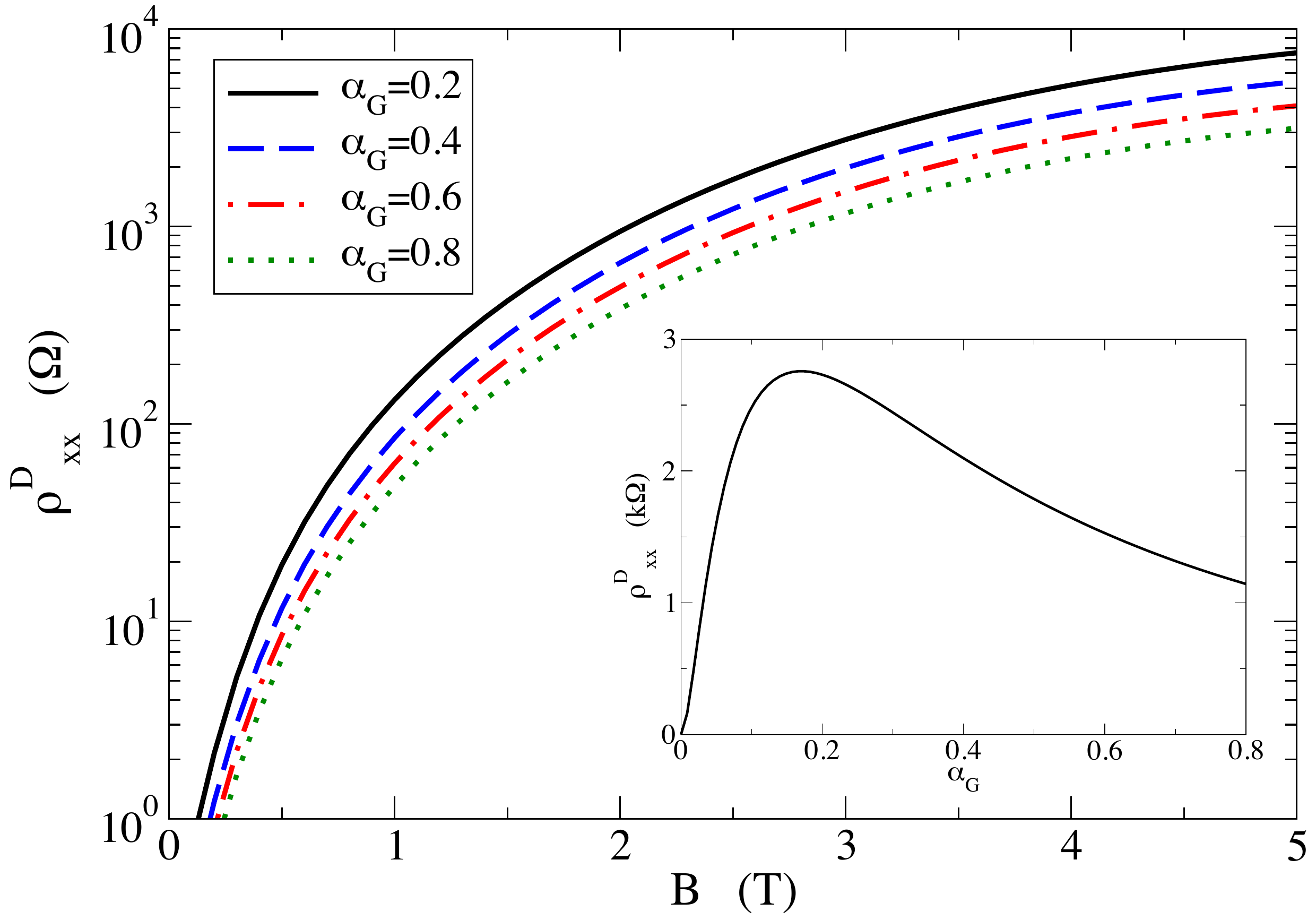}
\caption{(Color online)
Longitudinal drag resistivity at the simultaneous CNP $n_{\mathrm{L}} = n_{\mathrm{R}} = 0$ as a function of magnetic
field for different values of interaction coupling
$\alpha_{\mathrm{G}} = 0.2, 0.4, 0.6, 0.8$. Inset: The same quantity 
as a function of $\alpha_{\mathrm{G}}$ at a fixed magnetic field $B =
3\,\mathrm{T}$. $T$ and $d$ are the same as in previous figures.} 
 \label{rhoxx_B}
\end{figure}

Finally, we have calculated the longitudinal drag resistivity
$\rho_{xx}^{\mathrm{D}}$ at the simultaneous CNP as
a function of the magnetic field for different values of the interaction
parameter $\alpha_{\mathrm{G}}$. Fig.~\ref{rhoxx_B} shows that
$\rho_{xx}^{\mathrm{D}}$ is highly sensitive to changing magnetic
field strength, increasing to several $\mathrm{k}\Omega\mathrm{s}$
over a few teslas. It also shows that changing the
electron-electron interaction strength $\alpha_{\mathrm{G}}$ from
$0.2$ to $0.8$ counterintuitively 
\textit{decreases}
$\rho_{xx}^{\mathrm{D}}$. 
This 
finding can be explained by examining the $\alpha_{G}$  
dependence of the interlayer interaction $U(q,\omega,B)$ \cite{USupp}. 
Unlike the single layer case where the screened interaction 
$V \sim \alpha_{\mathrm{G}}/(1+\alpha_{\mathrm{G}})$ monotonically 
increases with $\alpha_{\mathrm{G}}$, the screened interlayer interaction of a bilayer 
$U \sim \alpha_{\mathrm{G}}/(1+\alpha_{\mathrm{G}}^2)$ decreases for large
$\alpha_{\mathrm{G}}$. This behavior is fully reflected in the drag 
resistivity as a function of $\alpha_{\mathrm{G}}$ depicted in the inset
of Fig.~\ref{rhoxx_B}. 

In summary, we find that the magneto-drag resistivity of graphene double layers
has a maximum and that the Hall drag resistivity vanishes at the simultaneous CNP.
The Hall drag resistivity is however comparable to the longitudinal resistivity 
at nearby densities, even though the Hall drag conductivity vanishes.  
Our theory accounts for momentum transfer due to interactions between
density-fluctuations in the two layers, but does not account for strong
correlations 
or address all
possible scenarios that have been raised \cite{Geim_drag2,Schutt} in connection 
with graphene double-layer magnetodrag.  The physics we explore must however 
contribute significantly to any magnetodrag measurements. 


Work at Alabama (WKT and JNH) was supported by startup and Research Grants Committee funds from
the University of Alabama.  AHM was supported by the DOE Division of Materials Sciences and 
Engineering under grant DE-FG02-ER45958 and by the Welch foundation under grant TBF1473.  BYKH acknowledges support from the
University of Akron for a Professional Development Leave.


\begin{thebibliography}{30}
\bibitem{eigen_remark} For compactness and uniformity, we use the same notation
  $\varepsilon_{s,n}$ and $\vert n, s, X
  \rangle$ for all LLs including $n = 0$, with  
$\varepsilon_{s,n = 0} \equiv 0$ and $\vert n = 0, s, X
  \rangle \equiv \vert 0, X \rangle$.
\bibitem{RPA} In strong magnetic fields, it is an intricate task to 
take into account the effect of screening properly covering a 
reasonably wide range of filling factors. The random phase
approximation (RPA) is commonly adopted \cite{Oppen,Nazarov} for the interlayer screened Coulomb potential
$U(q,\omega,B)$ to capture the main features of screening under strong magnetic fields in which fractional
quantum Hall effect is not yet developed.
\end{thebibliography}
\end{document}